\documentclass[sigconf]{acmart}

\AtBeginDocument{%
  \providecommand\BibTeX{{%
    \normalfont B\kern-0.5em{\scshape i\kern-0.25em b}\kern-0.8em\TeX}}}
    
\usepackage{xspace}
\usepackage{subcaption}
\usepackage{hhline}
\usepackage{hyperref}
\usepackage{enumitem}
\usepackage{color}  % add color to text
\usepackage{mathtools}

\usepackage{listings}
\usepackage[toc,acronym]{glossaries}

\newacronym{cbr}{CBR}{Constant Bit-Rate}
\newacronym{cod}{CoD}{Coefficient of Determination}
\newacronym{cqi}{CQI}{Channel Quality Indicator}
\newacronym{snr}{SNR}{Signal to Noise Ratio}
\newacronym{rsrp}{RSRP}{Reference Signal Received Power}
\newacronym{rsrq}{RSRQ}{Reference Signal Received Quality}
\newacronym{tp}{TP}{Throughput Prediction}
\newacronym{phy}{PHY}{physical}
\newacronym{ai}{AI}{Artificial Intelligence}
\newacronym{qoe}{QoE}{Quality of Experience}
\newacronym{qos}{QoS}{Quality of Service}
\newacronym{bs}{BS}{base station}
\newacronym{mldl}{ML/DL}{Machine and Deep Learning}
\newacronym{ml}{ML}{Machine Learning}
\newacronym{dl}{DL}{Deep Learning}
\newacronym{arima}{ARIMA}{Autoregressive Integrated Moving Average}
\newacronym{rtt}{RTT}{round-trip-time}
\newacronym{os}{OS}{Operating System}
\newacronym{api}{API}{Application Programming Interface}
\newacronym{rf}{RF}{Random Forest}
\newacronym{svm}{SVM}{Support Vector Machines}
\newacronym{wap}{WAP}{Wireless Acess Point}
\newacronym{lstm}{LSTM}{Long Short-Term  Memory}
\newacronym{rnn}{RNN}{Recurrent Neural Networks}
\newacronym{has}{HAS}{HTTP adaptive streaming}
\newacronym{ewma}{EWMA}{Exponential Weighted Moving Average}
\newacronym{ma}{MA}{Moving Average}
\newacronym{enodeb}{eNodeB}{evolved Node B}
\newacronym{aqm}{AQM}{Active Queue Management}
\newacronym{tcp}{TCP}{Transmission Control Protocol}
\newacronym{ip}{IP}{Internet Protocol}
\newacronym{rtmp}{RTMP}{Real-Time Messaging Protocol}
\newacronym{rtp}{RTP}{Real-Time Transport Protocol}
\newacronym{rtsp}{RTSP}{Real-Time Streaming Protocol}
\newacronym{rtcp}{RTCP}{Real-Time Control Protocol}
\newacronym{nat}{NAT}{Network Address Translation}
\newacronym{vod}{VoD}{Video on Demand}
\newacronym{udp}{UDP}{User Datagram Protocol}
\newacronym{quic}{QUIC}{Quick UDP Internet Connections}
\newacronym{cdn}{CDN}{Content Delivery Networks}
\newacronym{aimd}{AIMD}{Adaptive Increase Multiplicative Decrease}
\newacronym{mpc}{MPC}{Model Predictive Control}
\newacronym{dtx}{DTX}{Discontinuous Transmission Ratio}
\newacronym{rls}{RLS}{Recursive Least Squares}
\newacronym{sinr}{SINR}{Signal to Interference and Noise Ratio}
\newacronym{mcs}{MCS}{Modulation and Coding Scheme}
\newacronym{prb}{PRB}{Physical resource Block}
\newacronym{lte}{LTE}{Long-Term Evolution}
\newacronym{bdp}{BDP}{Bandwidth Delay Product}
\newacronym{fifo}{FIFO}{First In First Out}
\newacronym{vbr}{VBR}{Variable Bit-Rate}
\newacronym{red}{RED}{Random Early Detection}
\newacronym{wfq}{WFQ}{Weighted Fair Queuing}
\newacronym{sdn}{SDN}{Software-Defined Networking}
\newacronym{pq}{PQ}{Priority Queuing}
\newacronym{ecn}{ECN}{Explicit Congestion Notification}
\newacronym{dash}{DASH}{Dynamic Adaptive Streaming over HTTP}
\newacronym{hd}{HD}{High Definition}
\newacronym{gps}{GPS}{Global Positioning System}
\newacronym{ue}{UE}{User Equipment}
\newacronym{upplt}{UPPLT}{User Perceived Page Loading Time}
\newacronym{plt}{PLT}{Page Loading Time}
\newacronym{rssi}{RSSI}{Received Signal Strength Indicator}
\newacronym{re}{RE}{Resource Element}
\newacronym{vr}{VR}{Virtual Reality}
\newacronym{gan}{GAN}{Generative Adversarial Network}
\newacronym{soc}{SoC}{System on Chip}
\newacronym{bler}{BLER}{Block Error Rate}
\newacronym{acf}{ACF}{Autocorrelation Function}
\newacronym{ap}{AP}{Access Point}
\newacronym{adb}{ADB}{Android Debug Bridge}
\newacronym{hls}{HLS}{HTTP Live Streaming}
\newacronym{ran}{RAN}{Radio Access Network}
\newacronym{cv}{CV}{Cross-Validation}
\newacronym{cov}{CoV}{Coefficient of Variation}
\newacronym{cdf}{CDF}{Cumulative Distribution Function}
\newacronym{codel}{CoDel}{Controlled Delay Management}
\newacronym{fq}{FQ}{Flow Queue}
\newacronym{drr}{DRR}{Deficit Round Robin}
\newacronym{lr}{LR}{Latency Rate}
\newacronym{mtu}{MTU}{Maximum Transmission Unit}
\newacronym{sni}{SNI}{Server Name Identification}
\newacronym{http}{HTTP}{HyperText Transfer Protocol}
\newacronym{https}{HTTPS}{Hypertext Transfer Protocol Secure}
\newacronym{avc}{AVC}{Advanced Video Coding}
\newacronym{hevc}{HEVC}{High Efficiency Video Coding}
\newacronym{av1}{AV1}{AOMedia Video 1}
\newacronym{pid}{PID}{Proportional Integral Derivative}

\usepackage{listings} % to show code snippet
\usepackage{libertine}

\lstdefinestyle{BashInputStyle}{
  language=bash,
  basicstyle=\small\sffamily,
  numbers=left,
  numberstyle=\tiny,
  numbersep=3pt,
  frame=tb,
  columns=fullflexible,
  backgroundcolor=\color{gray!10},
  linewidth=0.95\linewidth,
  xleftmargin=0.1\linewidth
}

\newcommand{\ie}{{i.e.,}\xspace} 
\newcommand{\eg}{{e.g.,}\xspace}
\newcommand{\Kbps}{{kbps}\xspace}
\newcommand{\Mbps}{{Mbps}\xspace}

\lstdefinelanguage{Ini}
{
    basicstyle=\ttfamily\small,
    columns=fullflexible,
    tag=[s]{[]},
    tagstyle=\color{red}\bfseries,
    usekeywordsintag=true,
    morecomment=[l]{;},
    commentstyle=\color{gray}\ttfamily,
    alsoletter={=},
    ndkeywords={=},
    ndkeywordstyle=\color{green}\bfseries
}[html]

\copyrightyear{2022}
\acmYear{2022}
\setcopyright{acmcopyright}\acmConference[MMSys'22]{13th ACM Multimedia Systems Conference}{June 14--17, 2022}{Athlone, Ireland}
\acmBooktitle{13th ACM Multimedia Systems Conference (MMSys'20), June 14--17, 2022, Athlone, Ireland}
\acmPrice{15.00}
\acmDOI{xxx.xxxxxxxx}
\acmISBN{xxxxx.xxx.xxx}

\begin{document}

%%
%% The "title" command has an optional parameter,
%% allowing the author to define a "short title" to be used in page headers.
\title{Realistic Video Sequences for Subjective QoE Analysis}

%%
%% The "author" command and its associated commands are used to define
%% the authors and their affiliations.
%% Of note is the shared affiliation of the first two authors, and the
%% "authornote" and "authornotemark" commands
%% used to denote shared contribution to the research.

\author{Kerim Hodzic, Mirsad Cosovic, Sasa Mrdovic}
 \affiliation{%
 \institution{Faculty of Electrical Engineering, University of Sarajevo}
  \country{BiH}
 }

\email{{kerim.hodzic, mcosovic, smrdovic}@etf.unsa.ba}

\author{Jason J. Quinlan}
 \affiliation{%
 \institution{School of Computer Science \& Information Technology,\\University College Cork}
  \country{Ireland}
 }

\email{j.quinlan@cs.ucc.ie}

\author{Darijo Raca}
 \affiliation{%
\institution{Faculty of Electrical Engineering, University of Sarajevo}
 \country{BiH}
 }

\email{draca@etf.unsa.ba}

%%
%% By default, the full list of authors will be used in the page
%% headers. Often, this list is too long, and will overlap
%% other information printed in the page headers. This command allows
%% the author to define a more concise list
%% of authors' names for this purpose.

%%
%% The abstract is a short summary of the work to be presented in the
%% article.
\begin{abstract}
Multimedia streaming over the Internet (live and on demand) is the cornerstone of modern Internet carrying more than 60\% of all traffic. With such high demand, delivering outstanding user experience is a crucial and challenging task. To evaluate user \gls{qoe} many researchers deploy subjective quality assessments where participants watch and rate videos artificially infused with various temporal and spatial impairments. To aid current efforts in bridging the gap between the mapping of objective video \gls{qoe} metrics to user experience, we developed DashReStreamer, an open-source framework for re-creating adaptively streamed video in real networks. DashReStreamer utilises a log created by a \gls{has} algorithm run in an uncontrolled environment (\ie wired or wireless networks), encoding visual changes and stall events in one video file. These videos are applicable for subjective \gls{qoe} evaluation mimicking realistic network conditions.

To supplement DashReStreamer, we re-create 234 realistic video clips, based on video logs collected from real mobile and wireless networks. In addition our dataset contains both video logs with all decisions made by the \gls{has} algorithm and network bandwidth profile illustrating throughput distribution. We believe this dataset and framework will permit other researchers in their pursuit for the final frontier in understanding the impact of video \gls{qoe} dynamics.

%As live and on demand multimedia streaming over the internet becomes more and more popular, adaptive bitrate streaming techniques are implemented and getting improved to deliver the best Quality of Experience (QoE) to the users. To evaluate users QoE, subjective quality assessment where people watch and grade videos and objective quality assessment where videos are graded using one or many objective metrics are conducted. Videos being assessed are artificially distorted with startup delay, bitrate changes and stalls due to rebuffering events. To conduct more credible quality assessment, reproduction of original user experiences while watching different type of streams on different type and quality of networks is needed. For that purpose, we developed a cross-platform framework that can reproduce video from the log that contains information about video segments like segment index, bitrate, stall time and duration. Those logs can be created using internet bandwidth traces or logged on client while video is originally watched. We used bandwidth traces collected from real operational networks to create video logs and then used those video logs and framework to create dataset which is, to the best of our knowledge first publicly available dataset generated from real-word network bandwidth traces, and that contains video sequences, logs and bandwidth traces.

\end{abstract}

%%
%% The code below is generated by the tool at http://dl.acm.org/ccs.cfm.
%% Please copy and paste the code instead of the example below.
%%



\maketitle

\section{Introduction}
In its early days, the Internet was conceived with the idea for fast and reliable information sharing between many remote users. Since then, the Internet has transformed beyond basic e-mail communication, becoming one of the key pillars of modern society with multimedia entertainment at its heart, representing the dominant type of  traffic carried over today’s networks.

Video streaming depicts the main driver behind multimedia entertainment, accounting for almost 60\% of all Internet traffic in 2020. Furthermore, fuelled by a recent pandemic outbreak, forcing people to stay home, video traffic grew over the last two years, with applications such as YouTube, Netflix, Amazon Prime, Disney+ and Apple+ dominating overall traffic share~\cite{sandive2020}. 

The popularity of video streaming services led to user demand for high \gls{qoe} of delivered content. By definition, \gls{qoe} represents the magnitude of annoyance or the delight of a user's experience with an application or service~\cite{brunnstrom:hal-00977812}. However, measuring and modelling user \gls{qoe} is challenging due to its subjective intrinsic component. The challenge lies in modelling impairments that contribute to total \gls{qoe} score. These impairments include initial delay, stall events, average quality, switching frequency, and video duration~\cite{7222404}. Finding an optimal combination of these impairments to map to \gls{qoe} score is not a trivial task. Common approach includes performing subjective studies devising weights for each of the impairments~\cite{6573179,7222404, 7965631}. Many adaptive algorithms rely on these derived \gls{qoe} models, using them as an objective function in designing adaptation logic~\cite{10.1145/3458305.3463376, Yin:2015:CAD:2829988.2787486}. On the networks side, vendors usually rely on network metrics, such as packet loss and utilisation to map to user \gls{qoe}. 

The subjective evaluation of \gls{qoe} represents a foundation for better understanding and modelling user experience. Few studies perform both subjective and objective \gls{qoe} evaluation~\cite{duanmu2020assessing,8093636,5404314,6573179,7222404}. To estimate subjective experience, researchers design a few test sequences containing video impairments. Typically, these impairments are added artificially to the video sequence~\cite{7222404,10.1145/3458306.3458875}. However, in literature there are many datasets with bandwidth traces collected in various mobile environments under different wireless technologies~\cite{7218665,10.1145/2483977.2483991}. These datasets can be used for obtaining objective performance of adaptation algorithms including rate distribution, stall duration, and stall occurrence. Generating test video sequences based on realistic video logs complement the current literature on \gls{qoe}. To the best of our knowledge, there are no datasets generated based on real traffic patterns available to the research community. 

Motivated by this observation, we offer a framework for creating video sequences based on video logs collected either in real network or based on realistic bandwidth traces. Furthermore, we provide 234 video sequences based on video logs analysed over different bandwidth profiles collected from various wireless networks~\cite{9123139}. Video logs were generated by \gls{has} streaming algorithms under bandwidth profiles from different networks, resulting in a realistic snapshot of decisions algorithms made, including bitrate decisions (giving us rate distribution) and stall events (number and duration of stalls). 

In this paper, we present DashReStreamer~\footnote{\url{https://github.com/khodzic2/DashReStreamer}}, a framework for generating test video sequences with encoded stall and rate changes. In addition to the framework, we provide an extensive dataset containing video sequences created over 3G, 4G and WiFi networks. In total, 234 video sequences were generated with a duration of 5 minutes\footnote{\url{https://shorturl.at/dtISV}}. The dataset contains video logs and bandwidth traces used for the video sequence generation. These video sequences are suitable for subjective \gls{qoe} evaluation, and can aid in the better understanding of user experience in different scenarios. To the best of our knowledge, our \gls{qoe} dataset is the first publicly available dataset that contains video sequence, logs, and bandwidth traces. 

The remainder of this paper is organised as follows. Section~\ref{sec:backgroundDataset} describes related work regarding similar datasets and \gls{qoe}-related video metrics. The overview and key features of proposed framework are explained in Section~\ref{sec:framework_overview}, while Section~\ref{sec:dataset_overview} provides an overview of the dataset generated by DashReStreamer. In Section~\ref{sec:future_work} we layout future work, while Section~\ref{sec:conclusion} outlines our conclusion.

%\noindent
%\textbf{Dataset and framework are available at \url{https://shorturl.at/dtISV}\footnote{\url{https://drive.google.com/drive/folders/1zIwED62FZYp52cMCD_OFeziuzUeiGZ4x?usp=sharing}}}

\section{Background and Related Work}
\label{sec:backgroundDataset}
The main goal of \gls{has} algorithms is maximising user perceived QoE. This daunting task relies on accurate representation of subjective impact through mapping objective \gls{qos} metrics at client side (\eg initial delay, average bitrate, re-buffering events, and switching frequency) or metrics measured at the network such as utilisation and packet-loss rate. Also, the majority of proposed \gls{has} algorithms in literature relies on \gls{qoe} models to quantitatively compare its performance to existing state-of-the art \gls{has} algorithms. Furthermore, \gls{qoe} models expressed as linear combination of impairments ~\eqref{eq:qoe_eq}, represent a suitable candidate for designing a \gls{has} algorithm that maximises a given \gls{qoe} model. A typical approach includes modelling the \gls{qoe} model as the utility function of the optimisation problem~\cite{10.1145/3204949.3204961, 10.1145/3458305.3463376, Yin:2015:CAD:2829988.2787486}.

A typical template equation used for deriving \gls{qoe} model is~\cite{6573179,7222404,7965631}:
\begin{equation}
\label{eq:qoe_eq}
\begin{split}
    \text{QoE}_s = w_o \cdot \text{QoE}_m - (w_{t} \cdot I_{t} + w_{v} \cdot I_{v} )+ f(I_{t}, I_{v}),
\end{split}
\end{equation}
where $I_{t}$ represents temporal impairment factor, and $w_{t}$ represents its weight. Temporal quality impairments indicate degradation due to initial delay and stall performance (stall number and stall duration). While initial delay has a minor negative effect on \gls{qoe} (up to 16 seconds), stall events have the highest negative impact on overall user experience~\cite{6913491}. $I_{v}$, and $w_{v}$ represent visual quality impairment factors and its weight, respectively. 

Average bitrate and switching behaviour model visual quality impairments. Similar to stall performance, bitrate quality amplitude has a significant effect on \gls{qoe}~\cite{6982305}, unlike switching between different qualities while retaining the same resolution~\cite{6982305}. However, switching between different resolutions can influence user experience~\cite{8019297}. $\text{QoE}_m$ depicts the maximum (initial) value (score) for \gls{qoe} or growth factor depending on the \gls{qoe} model, and $w_o$ denotes a weight for the $\text{QoE}_m$ score. Some \gls{qoe} models take into account impairments that occur simultaneously. In these scenarios, aggregate subjective effect is not a direct sum of each impairment~\cite{7222404}. The role of function $f(I_{t}, I_{v})$ is to compensate for this effect.

However, these impairments (\ie metrics) are mutually contradictory. High bitrate increases the chance of buffer underflow resulting in stall events, while streaming at low bitrate quality has a severe negative impact on perceived user experience. 

To capture the mapping between user perceived experience and objective metrics, many studies use subjective evaluation. This evaluation relies on assessing video quality by participants in a controlled lab environment~\cite{7222404, 7965631, 7879860, shang2021assessment}. Each participant rates a video sequence on a 100-point scale (denoted as $R$, where some studies use 5 or 10-point scale). The procedure is repeated for a series of test sequences. Each test sequence is embellished with one or more impairments. Finally, for each test sequence and given score $\text{R}$, the impairment impact is calculated as $\text{100-R}$. 

Subjective evaluation is an expensive, time-consuming process performed with a limited number of human subjects (usually around 30) restricting the statistical validity of collected results. Alternatively, some studies opt for a crowd-sourcing approach, where a large number of users rate video sequences online in an uncontrolled environment~\cite{10.1145/3458306.3458875, 10.1145/2398776.2398799, 6573179}.

The main challenge for subjective evaluation is augmentation of the test video sequences with particular impairments. Typically, these impairments are artificially created and added to video clips. However, artificially created impairments do not necessarily reflect impairments observed in real network conditions, either their frequency (\eg number of rate switches, number of stalls), or duration (\eg stall duration).

There are a plethora of bandwidth datasets collected in real networks available in literature~~\cite{7218665,10.1145/2483977.2483991, 10.1145/3204949.3208123}. These datasets reflects real conditions observed in networks and can be leveraged for realistic creation of temporal and visual impairments.

Motivated by the lack of video sequences with the impairments based on real network conditions, we designed a tool for creating video sequences with impairments collected from video sessions collected over realistic bandwidth traces. We believe this dataset will aid in ongoing research to better understanding factors affecting user experience.

\section{D\lowercase{ash}R\lowercase{e}S\lowercase{treamer} Overview}
\label{sec:framework_overview}

DashReStreamer provides the functionality to reproduce network impact on video player performances by creating video clips including all resolution changes and re-buffering events. We achieve this functionality by utilising video logs generated by the client during the original stream of content in an uncontrolled environment (\ie real production network).

Typically these logs include various information related to \gls{has} \gls{qos} metrics (\eg bitrate, switches and stall information). To illustrate, \tablename~\ref{tab:godash_output_log_headers} depicts an example of a video log.

{\renewcommand{\arraystretch}{1.2}
\setlength\tabcolsep{1.5pt}
\begin{table}[htb]%
	%\footnotesize
	\centering
	\caption{Sample output from the video log}
	\label{tab:godash_output_log_headers}
	\begin{tabular}{p{20mm}  p{47mm} p{10mm}}
		\textbf{Type}   & \textbf{Description} & \textbf{Unit}\\
		    \hline
		        Seg\_\# &       Streamed segment number &- \\
                %Seg\_\# &  Streamed segment number (full, main, live)\\ 
                %& Streamed segment range (full\_byte\_range,  main\_byte\_range)\\
                Arr\_Time &     Arrival time & ms\\
                Del\_Time &     Time taken to receive the segment & ms\\
                Stall\_Dur &    Stall duration & ms\\
                Rep\_Level &    Representation Quality & kbps\\
                Del\_Rate &     Delivery rate & kbps \\
                Act\_Rate &     Actual rate & kbps \\
                Byte\_Size &    Size of segment& byte\\
                Buffer\_Level &        Buffer level & ms\\
                
		\hline
	\end{tabular}
    %\vspace{-0.1in}
\end{table}
}
From the DashReStreamer perspective, three features are necessary for video clips generation.
These features are details on:
\begin{itemize}
\item Segment number.
\item Segment bitrate: we use this information to select the subsets of downloaded segments during playback (currently DashReStreamer does not support byte-range \gls{has} content).
\item Stall events: we use stall events (occurrence and duration) to add stalls (\eg duplicating last frame of segment) at the end of segments affected by re-buffering events. 
\end{itemize}

\subsection{Framework Implementation}

We use the Python programming language and FFmpeg\footnote{\url{https://www.ffmpeg.org/}} library for the implementation of DashReStreamer. FFmpeg is a cross-platform multimedia framework for transforming (\ie encoding, decoding, transcoding, mux, demux, stream, and filter) a wide range of media formats (video and image).  

DashReStreamer starts by parsing video log file, where used bitrates of video segments are identified for further processing. Two methods are used for this process:
\begin{itemize}
\item \textit{read\_replevels\_log} - takes four arguments (file path, index column name, bitrate column name, and type of separator, e.g., csv), parses the file and returns the output as a hash function (dictionary) storing the bitrate for each downloaded segment.
\item \textit{read\_stalls\_log} - takes four arguments and returns the position and duration of each stall, where position is related to segment when the stall happened.
\end{itemize}
The next step includes filtering a subset of streamed segments. Segments can be stored locally or remotely on a web server. In the latter case, an mpd file is used for downloading the streamed segments from the server to the local machine.

Three methods are used for the manipulation of the streamed segments (for the case when segments are already stored locally):
\begin{itemize}
\item \textit{copy\_init\_file} - takes two arguments, location of init mp4 file and destination where init file will be stored for further processing.
\item \textit{copy\_video\_segments} and  \textit{copy\_audio\_segments} - are methods for copying downloaded segments for further processing. Similar to \textit{copy\_init\_file} method, these methods take two arguments. These methods rely on a dictionary created in the previous step by the \textit{read\_replevels\_log} and \textit{read\_stalls\_log} methods for appropriate identification of streamed segments and stall events.
\end{itemize}

For the case when segments are directly downloaded from a web server, the script uses the location of the mpd (\ie URL) to retrieve only the subset of segments streamed in the video logs. This procedure is similar to the behaviour of traditional \gls{has} client~\cite{10.1145/1943552.1943572} (without actual decoding of the data). We use an existing library for parsing mpd files\footnote{\url{https://github.com/sangwonl/python-mpegdash}}. There are three main methods for preparing the data linking in this step:
\begin{itemize}
\item \textit{parse\_mpd} - the method that parses mpd files and stores urls of audio and video segments into a dictionary,
\item \textit{download\_video\_segments} - takes two arguments, location of mpd file and destination folder,
\item \textit{download\_audio\_segments} - similar to the previous method, method downloads audio segments.
\end{itemize}

DashReStreamer proceeds by combining segments with init file (originally segments are in m4s format). For this operation we use two methods:
\begin{itemize}
\item \textit{prepare\_video\_init} - takes two arguments, location of video segments and init file,
\item \textit{prepare\_audio\_init} - prepares audio segments similar to the previous method.
\end{itemize}

The output of these methods are new audio and video segments (in avi\footnote{Audio Video Interleave} and mkv\footnote{Matroska Multimedia Container} format respectively) which can be played independently. Next, we combine the individual pairs of audio and video segments, using the FFmpeg library. This operation is performed by method \textit{concat\_audio\_video\_ffmpeg} which takes two arguments: location of segments and flag indicating should segment be rescaled to different resolution. 

We create a video sequence combining segments including all bitrate/resolution changes and stall events. First, we create stall-induced segments. For the creation of stall-induced segments, we take the stall duration and the segment just before stall starts. We take the last frame of the identified segment, and add them at the end of segment for the duration of the stall. Finally, we add gif\footnote{Graphics Interchange Format} as an overlay on top of the stall-induced segments. After all segments are prepared, we join them into a final mkv video file. The preceding logic is implemented in method \textit{concat\_audio\_video\_ffmpeg\_final}. This method takes three arguments: location of segments, location of gif and destination for final video.

\subsection{Example of Use}
There are several options available to run DashReStreamer, either directly through the command line or using a configuration file. For command line use, \tablename~\ref{tab:qoe_framework_arguments} depicts the supported options for running the framework. 

{\renewcommand{\arraystretch}{1.2}
\setlength\tabcolsep{1.5pt}
\begin{table}[htb]%
	%\footnotesize
	\centering
	\caption{Options for running QoE framework}
	\label{tab:qoe_framework_arguments}
	\begin{tabular}{p{20mm}  p{60mm} }
		\textbf{Parameter}   & \textbf{Description}\\
		    \hline
		  path\_to\_log & Location of video log \\
          rep\_lvl\_col &  Column name used in video log for bitrate\\
          seg\_index\_col &  Column name used in video log for segment index\\
          stall\_dur\_col & Column name used in video log for stall duration\\
          log\_separator &  Separator used in video log (example: tab)\\
          config\_path & Location of config file\\
          path\_video & Location of video segments\\
          path\_audio & Location of audio segments\\
          gif\_path & Location of gif file\\
          log\_location & Flag indicating location of segments (local or remote)\\
          dest\_video & Location where to save intermediate files during processing (segments)\\
          final\_path & Location where final concated video is saved\\
          parameter\_type & Flag indicating use of command line arguments or config file\\
          cleanup & Flag indicating removal of intermediate files (segments)\\
          auto\_scale & Options for enabling auto-scaling of segment resolution \\
          scale\_res & Rescaling segments to predetermined resolution (example: 1080p)\\

		\hline
	\end{tabular}
    %\vspace{-0.1in}
\end{table}}

\noindent
\textbf{Case \#1:} For segment files stored locally, the command outlined in Listing~\ref{lst:framework-terminal} produces a video file based on the video log file.

\begin{lstlisting}[style=BashInputStyle, 
     captionpos=b, caption={Example of creating video from local segments},
     label={lst:framework-terminal}]
# python video_log_merger.py --path_to_log video_log.log 
--rep_lvl_col Rep_Level 
--seg_index_col Chunk_Index 
--log_separator tab 
--stall_dur_col Stall_Dur 
--path_video ./sintel/DASH_Files/full/ 
--dest_video ./tmp_files/ 
--path_audio ./sintel/DASH_Files/audio/full/ 
--gif_path ./gif.gif 
--final_path ./final/ --parameter_type path
--cleanup True
\end{lstlisting}
The depicted example in in Listing~\ref{lst:framework-terminal} utilises the open-source movie Sintel, filters segment qualities used by adaptation algorithm outlined by video log file (video\_log.log file), re-creates video sequence adding stall events (with the re-buffering image) and saves the output to the folder final. This command retains native resolution for each segment causing a visual change in aspect ratio when segments of the video switch from one resolution to another. 

Alternatively, we can mandate that all segments have the same output resolution through the option of autoscaling. We support two types of autoscaling: scaling to the highest resolution observed in the log file, or scaling to predetermined resolution given by parameter \textit{scale\_res}. The Listing~\ref{lst:framework-fixed-resolution} example shows how to create an output video file with a fixed 1080p resolution for all segments. 

\noindent
\textbf{Case \#2:} Creating video file with same predetermined resolution is depicted in Listing~\ref{lst:framework-fixed-resolution}
\begin{lstlisting}[style=BashInputStyle, 
     captionpos=b, caption={Example of creating video with same resolution for all segments},
     label={lst:framework-fixed-resolution}]
# python video_log_merger.py --path_to_log video_log.log 
--rep_lvl_col Rep_Level 
--seg_index_col Chunk_Index 
--log_separator tab 
--stall_dur_col Stall_Dur 
--path_video ./sintel/DASH_Files/full/ 
--dest_video ./tmp_files/ 
--path_audio ./sintel/DASH_Files/audio/full/ 
--gif_path ./gif.gif 
--final_path ./final/ --parameter_type path
--scale_resolution 1080p 
--auto_scale 2 
--cleanup True
\end{lstlisting}
Similar to Listing~\ref{lst:framework-terminal}, we recreate an output video clip from the video log file, with the difference that we scale each segment to a Full HD resolution. This option is achieved by setting auto\_scale to 2 (where we have three supported values 0,1,2), and setting scale\_res to 1080p.

The DASHReStreamer framework also supports the use of a configuration file as input to the python script.   Listing~\ref{lst:config} illustrates an example configuration file. Note that all the input parameters are the same as the parameters used for the command-line input.

\begin{lstlisting}[language={Ini}, captionpos=b,caption={Example of config file}, label={lst:config}]
    [parameters]
    parameters = config
    path_to_log = <path>
    rep_lvl_column = Rep_Level
    chunk_index_column = Chunk_Index
    stall_dur_column = Stall_Dur
    log_separator = tab
    path_audio = <path to audio segments>
    path_video = <path to video segments>
    dest_video = <where to save/download segments>
    gif_path = <path to gif file>
    final_path = <where to save final video>
    mpd_path = <url for mpd file>
    auto_scale = 0
    log_location = local
    \end{lstlisting}

\section{QoE dataset overview}
\label{sec:dataset_overview}
This section gives a short overview of the dataset used for generating various video sequences in different wireless conditions. The majority of the video sequences contain at least one re-buffering event as those cases are the most interesting for \gls{qoe} modelling.

\subsection{Video Logs Generation}
\begin{figure}[htb]
	\centering
	\includegraphics[width=0.8\columnwidth]{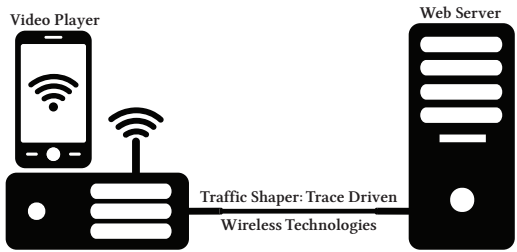}
	\caption{The data-driven generation testbed.}
	\label{fig:generalized_testbed}
	\vspace{-0.1in}
\end{figure}

We use video logs generated by experiments in~\cite{9123139} for the creation of the video sequences. The video logs are generated based on bandwidth traces collected from real operational networks. \figurename~\ref{fig:generalized_testbed} illustrates a generalised testbed used for producing video logs.

The testbed consists of a server machine, an intermediate device, \eg \gls{wap} and one or more wireless-capable end devices (\ie mobile device). The server machine performs two roles, one  as web server for video content, and second as traffic shaper for link between server and intermediate device. The traffic shaping procedure includes the use of traffic shaping tools like Linux traffic control (tc) for emulation of different bandwidth profiles from collected bandwidth logs. The intermediate device connects to the end device via WiFi channel. Finally, the end device streams content from the server via a bottleneck link creating the video log after streaming finishes.

A 4K encoded animation clip is used as the video content stored at the server. The clip is encoded, using the H.264/AVC codec, into thirteen bitrates (from 235\Kbps to 40\Mbps) across eight resolutions.

For traffic shaping, bandwidth logs were used from three different wireless technologies, 3G, 4G and WiFi, including various mobility patterns (static, pedestrian, bus, car and tram). \tablename~\ref{tab:stats_wireless_technologies} shows summary statistics (average and standard deviation of measured bandwidth) for these logs~\cite{9123139}.

\begin{table}[ht]%
	\small
	\centering
	\caption{Throughput Statistics for collected bandwidth logs}
	\label{tab:stats_wireless_technologies}
	\begin{tabular}{c|c|c}
        \textbf{Technology} & \textbf{Average (Mbps)} & \textbf{Standard Deviation (Mbps)}\\\hline
        3G&1.26&0.97\\
        4G&11.32&13.17\\
        WiFi&18.71&17.73\\
		\hline
	\end{tabular}
\vspace{-0.1in}
\end{table}
%Four state-of-the-art adaption algorithms were used for content streaming, including Arbiter+~\cite{8334618}, BBA2~\cite{10.1145/2740070.2626296}, BOLA~\cite{9110784}, and Elastic~\cite{6691442}.

\subsection{Video Sequences Generation}
We utilise video logs explained in the previous section and our proposed tool (see Section~\ref{sec:framework_overview}) to generate 234 video impaired clips.
For video content, we select three open-source clips from~\cite{10.1145/3204949.3208130}. These clips are Big Buck Bunny (BBB)\footnote{\url{https://peach.blender.org/}}, Sintel\footnote{\url{https://durian.blender.org/about/}}, and Tears of Steel (TOS)\footnote{\url{https://mango.blender.org/about/}}. 
Big Buck Bunny is an animation clip with a duration of 10 minutes and 34 seconds. The content is composed of animated characters with a non intricate background, encoded with a maximum 4K resolution of 3840x2160, at 60fps. Similarly, Sintel is an animation clip with a duration of 14 minutes and 48 seconds. The content is composed of complex animated characters and scenery, encoded with a maximum 4K resolution of 3840x2160, at 24fps. Finally, Tears of Steel is a movie-alike clip enhanced with digital visual effects of 12 minutes and 14 seconds duration. The content is composed of real actors and superimposed digital effects, encoded with maximum 4K resolution of 3840x2160, at 24fps.

All clips are encoded in thirteen bitrates and eight different resolutions as depicted in \tablename~\ref{tab:clip_resolution} and sourced from~\cite{10.1145/3204949.3208130}. Also, all clips are encoded with the sound of five minutes plus total stall duration. We select 27, 25, and  26 video logs generated from 3G, 4G, and WiFi network traces, respectively. \tablename~\ref{tab:video_qos_stats} depicts video \gls{qos} metric statistics for the selected logs.

\begin{table}[htbp]
\centering
      \caption{Ladder for the average encoding rate, and resolution for the used dataset}
      \label{tab:clip_resolution}
      %¥begin{tabular}{clclclclclclclclclclclclcl}
      	\begin{tabular}{lll}
       	 	\toprule
       	 	\textbf{No.} & \textbf{Bitrate} & \textbf{Resolution} \\
       	 		 13 &40\Mbps & 3840x2160  \\
       	 		 12 &25\Mbps & 3840x2160  \\
       	 		 11 &15\Mbps & 3840x2160  \\
         		 10 &4.3\Mbps & 1920x1080  \\
         		 9 &3.85\Mbps & 1920x1080  \\
         		 8 &3\Mbps & 1280x582  \\
         		 7 &2.35\Mbps & 1280x582  \\
         		 6 &1.75\Mbps & 720x328  \\
         		 5 &1.05\Mbps & 640x292  \\
         	     4 &750\Kbps & 512x234  \\
         		 3 &560\Kbps & 512x234  \\
         		 2 &375\Kbps & 384x174  \\
         		 1 &235\Kbps & 320x146  \\
        		\midrule
      	\end{tabular}
    \end{table}

\begin{table*}[htbp]
\centering
      \caption{Average \gls{qos} metrics for selected video logs}
      \label{tab:video_qos_stats}
      %¥begin{tabular}{clclclclclclclclclclclclcl}
      	\begin{tabular}{ccccc}
       	 	\toprule
       	 	\textbf{Network} & \textbf{Bitrate (Mbps)}  & \textbf{Num. Switches} &  \textbf{Num. Stalls} & \textbf{Stall Dur. (s)}  \\
       	 		 3G &1.6 & 19.6 & 3.4 & 53.9 \\  
       	 		 4G &5.8 & 18.8 & 0.96 & 14.3 \\  
       	 		 WiFi &6.3 & 12.5 & 0.77 & 1.95  \\
        		\midrule
      	\end{tabular}
    	
\end{table*}
    
Video logs from 3G network traces have the highest number of stalls and stall duration followed by 4G and WiFi network traces. This result is intuitive as indicated by throughput statistics in \tablename~\ref{tab:stats_wireless_technologies}. Also, WiFi network traces are mostly collected in a static environment thus having the highest average throughput.

\section{Future Work}
\label{sec:future_work}
While our framework currently offers a mechanism to generate an adaptive video dataset, which can be used in subjective testing or similar research settings, typically using a five-point MOS scale, future work will include the calculation of Video Quality Metrics such as PSNR, SSIM, VMAF and P.1203~\cite{7965631} for each generated clip.
Creating additional KPIs through which video QoE and Network QoS can be determined.

Furthermore DashReStreamer currently only supports the \textit{full} profile of the DASH standard. Future work includes adding support for remaining mpd profiles (\ie main, live, onDemand, and byte range) and other \gls{has} datasets available in the literature. We also plan on adding realistic video clips for different \gls{has} segment durations to our Dataset, including segment durations of between 2 and 10 seconds, allowing for a much richer and diversified \gls{qoe} video dataset. 
% % \input{future_work}
% \input{use_case}
% \input{conclusion}
\section{Conclusions}
\label{sec:conclusion}
In this paper, we present DashReStreamer, an open-source cross-platform framework for reproducing adaptively streamed video from real operational networks. DashReStreamer allows re-creating video clips with all bitrate/quality changes and stall events. Generated video clips mimic decisions made by \gls{has} adaptation algorithms, and the selected bitrates chosen under realistic time-varying conditions observed in the network. The framework utilises video logs produced by \gls{has} adaptation algorithm to re-create video clips. Furthermore we generate 234 video clips mimicking the behaviour of various \gls{has} adaptation algorithms under three different wireless technologies (\ie 3G, 4G, and WiFi), producing a dataset with realistic bitrate changes and stall events.  We believe this dataset will help researchers in better understanding factors affecting user experience for \gls{has} multimedia technologies, aiding its use in both objective and subjective \gls{qoe} evaluation.

\begin{acks}
The authors acknowledge the support of the Ministry of Education, Science and Youth of Sarajevo Canton.
\end{acks}

%%
%% The next two lines define the bibliography style to be used, and
%% the bibliography file.

\bibliographystyle{ACM-Reference-Format}
\bibliography{lit.bib}

\end{document}